\documentclass[a4paper,fleqn,usenatbib]{mnras}

\usepackage{mathptmx}

\usepackage[T1]{fontenc}
\usepackage{ae,aecompl}

\usepackage{graphicx}	
\usepackage{amsmath}	
\usepackage{amssymb}	

\newcommand{\sr}{2FHL J1944.9$-$2144}
\newcommand{\mss}{MST 0757$-$3728~}
\newcommand{\ms}{MST 0757$-$3728}
\newcommand{\mfss}{MST 1945$-$2143~}
\newcommand{\mfs}{MST 1945$-$2143}

\title[Two new high energy $\gamma$-ray blazars]{Two new high energy $\gamma$-ray blazar candidates}

\author[R. Campana et al.]{
R. Campana,$^{1}$\thanks{E-mail: campana@iasfbo.inaf.it (RC)}
A. Maselli,$^{2}$
E. Bernieri$^{3}$
and E. Massaro$^{4,5}$
\\
$^{1}$INAF/IASF-Bologna, via Piero Gobetti 101, I-40129 Bologna, Italy\\
$^{2}$INAF/IASF-Palermo, via Ugo La Malfa 153, I-90146 Palermo, Italy\\
$^{3}$INFN-Sezione di Roma Tre, via della Vasca Navale 84, I-00146 Roma, Italy\\
$^{4}$INAF/IAPS, via Fosso del Cavaliere 100, I-00133 Roma, Italy\\
$^{5}$In Unam Sapientiam, Piazzale A. Moro 2, I-00185 Roma, Italy\\
}

\date{Accepted XXX. Received YYY; in original form ZZZ}

\pubyear{2016}

\begin{document}
\label{firstpage}
\pagerange{\pageref{firstpage}--\pageref{lastpage}}
\maketitle

%
\begin{abstract}
We report the detection of two new $\gamma$-ray sources in the Fermi-LAT sky (Pass 8) at energies higher than 20~GeV, and confirmed at lower energies, using a source detection tool based on the Minimum Spanning Tree algorithm. One of these sources, at a Galactic latitude of about $-4$\degr, is a new discovery, while the other was previously reported above 50~GeV in the 2FHL catalogue. 
We searched for archival multi-wavelength data of possible counterparts and found interesting candidates.
Both objects are radio sources and their WISE infrared colours are typical of blazars. While for the former source no optical spectra are available, for the latter a puzzling optical spectrum corresponding to a white dwarf star is found in the 6dF database. We discuss the spectral energy distributions of both sources and possible interpretations.
\end{abstract}



\section{Introduction}

The entire $\gamma$-ray sky in the 0.03 to $>$300 GeV band is continuously observed
by the \textit{Fermi}-Large Area Telescope (LAT) experiment \citep{atwood09} since 
August 2008.
To date, the Fermi collaboration has published three catalogues of sources detected above
100~MeV, the most recent of which \citep[3FGL catalogue,][]{3FGL}
is based on the first 4 years of data, and two catalogues of high energy sources above 
10~GeV \citep[1FHL,][]{1FHL}
and 50~GeV \citep[2FHL,][]{2FHL},
the latter covering a time interval of 80 months.
In addition, some specific catalogues for selected classes of objects such as pulsars 
\citep{abdo13}
and Supernova Remnants \citep{acero16}
have also been published.
The continuous increase of observational data, 
in addition to the improved new data processing and instrumental response functions (Pass 8)
has enriched the quality of the \textit{Fermi}-LAT $\gamma$-ray sky and has raised its capability
to detect new sources.

In a series of papers we successfully applied a tool based on the Minimum Spanning Tree (MST) algorithm 
\citep{campana08, campana13} to search for photon clusters in the $\gamma$-ray sky at
Galactic latitudes higher than 25\degr, and to associate them with known blazars or candidates 
within an angular distance of 6\arcmin\ \citep{paperI,paperII,paperIII,paperIV}.
In the present paper we report the detection, with the same MST clustering method, of 
two new high energy $\gamma$-ray sources found while analyzing the \textit{Fermi}-LAT 
Pass 8 sky in regions below the previous Galactic latitude limit.
The new sources were first detected applying the MST analysis at energies higher
than 20 GeV and then confirmed by similar analysis at lower energies.
One of these clusters was found to coincide with the source 2FHL J1944.9$-$2144 previously 
detected above 50 GeV \citep{2FHL}, but never reported at lower energies.

We completed the analysis of these two sources by means of the \emph{Fermi} Science 
Tools\footnote{\url{http://fermi.gsfc.gov/ssc/data/analysis/software}}, based on the Maximum 
Likelihood (ML) method, to obtain 
count maps, light curves and Test Statistics ($TS$) values. 
As a final step, we searched in data archives and in the literature for possible counterparts at lower energies, 
from the radio to the soft X rays, and discussed their properties to confirm their possible blazar nature, 
as well as some open issues for one of them.

\section{MST detection of $\gamma$-ray clusters}
MST is a topometric cluster-finding algorithm that exploits the pattern of ``connectedness'' 
of the detected photons, treated as the \emph{nodes} in a graph, where the \emph{edges} 
are the angular distances connecting them.
The advantage of MST 
is the capability to quickly find potential $\gamma$-ray sources by examining only the 
incoming directions of the photons, regardless of their energy distribution.
In the specific application of MST to the $\gamma$-ray sky we defined (and successfully 
tested on simulated and real fields) various selection parameters, useful to assess the 
significance of the detected clusters and therefore their possible nature as genuine 
astrophysical sources \citep{campana08,massaro09,campana13}.
In particular, we found that the \emph{cluster magnitude} $M$ is a very good 
indicator of the detection significance; it is defined as  $M = ng$, where
$n$ is the number of photons in a cluster, and $g$ is the  \emph{clustering degree}, i.e. the 
ratio of the mean edge length in the cluster to the mean value $\Lambda_m$ in the field.
As shown by \citet{campana13}, in regions with a Galactic latitude high enough to have a 
rather uniform and not very dense background, $M$ values higher than 20 correspond to 
significance values close to and generally higher than 4 standard deviations.
The centroid coordinates, obtained by computing a weighted mean of the photons' coordinates 
\citep[see for details][]{campana13}, give an indication if a cluster is compatible with a $\gamma$-ray source.

\begin{figure*}
\centering
\includegraphics[width=\textwidth]{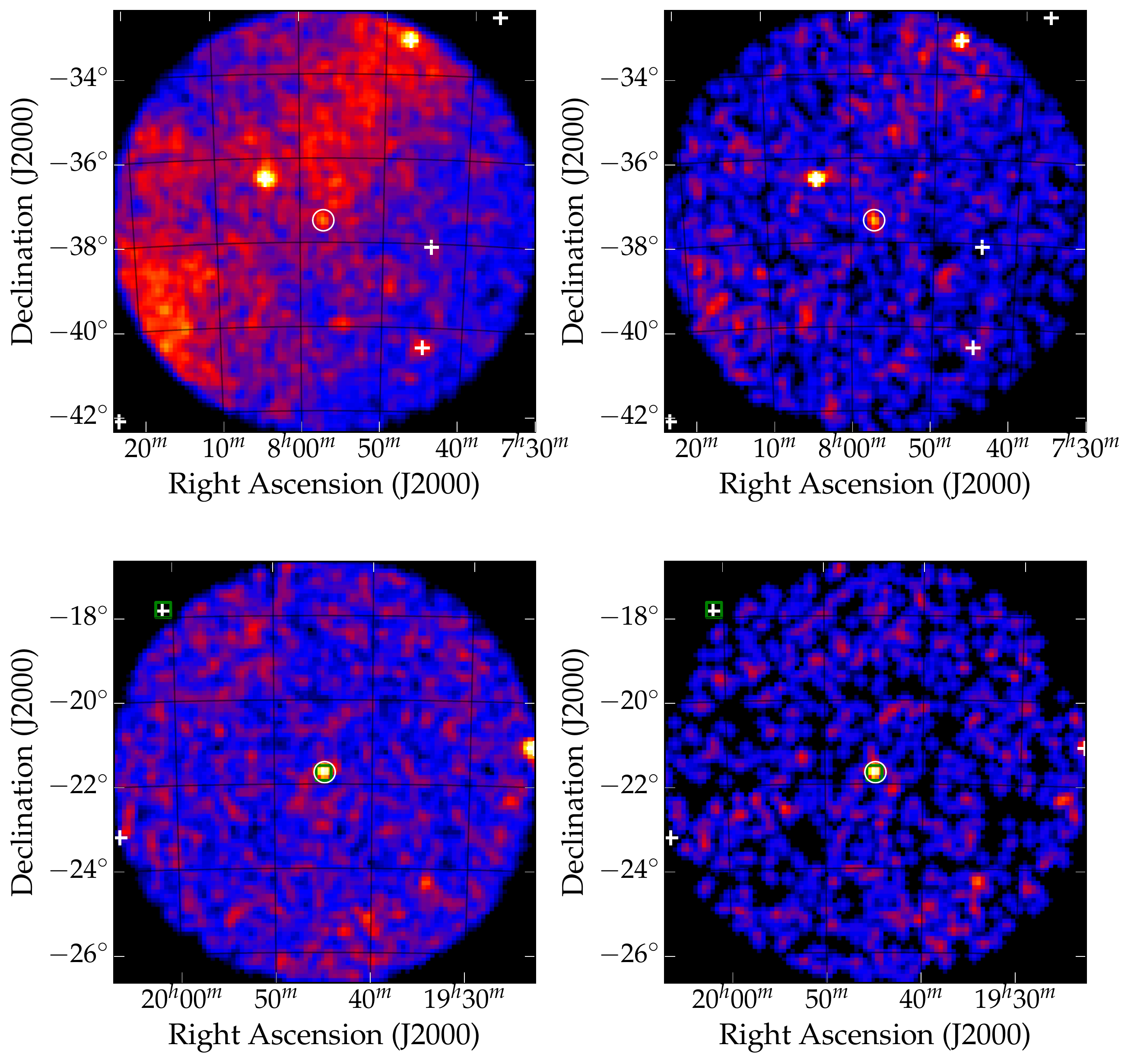}
\caption{
Upper panels:
False colour count images in equatorial coordinates at energies higher than 
3 GeV (left) and 10 GeV (right) of the sky region centred at the 
cluster \ms. 
Lower panels:
The same for the sky region centred at the 
cluster \sr.
For all the count maps the radius of the considered circular region is 5\degr.
White cross and green square symbols indicate the positions of 3FGL and 2FHL sources, respectively, 
while the white circles correspond to the MST cluster centroid coordinates.
}
\label{f:fig1}
\end{figure*}

Once detected with the MST algorithm, the clusters can be further analysed with other well recognized statistical methods, 
such as the maximum likelihood \citep[ML,][]{mattox96}, in order to obtain an independent evaluation 
of their statistical significance and to study the time and energy properties of the corresponding source.
In this sense, the MST method has been already applied to obtain lists of seed clusters for the 1FGL and 2FGL 
\textit{Fermi}-LAT catalogues \citep{abdo10a, nolan12}.

In the present work we have considered all the data collected in 7 years by \textit{Fermi}-LAT, from 2008 
August 04 to 2015 August 04, both in a region close to the Galactic equator not distant from the
very bright Vela pulsar, and in two intermediate Galactic latitude belts ($14\degr < |b| < 26\degr$).
Pass 8 data were filtered using the standard LAT Science Tools routines \texttt{gtselect} and 
\texttt{gtmktime} with standard cuts on the zenith angle and on 
the rocking angle to limit contamination from the bright Earth limb.

\begin{table*} 
\caption{Coordinates and main properties of MST clusters detected at energies
higher than 20~GeV in the first considered region.
The first pair of columns gives the J2000 coordinates of the MST clusters, while the second pair
the Galactic coordinates. The other columns give the number of photons, the clustering factor,
the magnitude, the median and the maximum radii (i.e. the radii from the cluster centroid coordinates 
that contain 50\% and all the cluster photons, respectively), and the angular separation from the 3FGL source.
The last two rows report the cluster parameters detected in a narrower field but with $E>10$ GeV and
$E>3$ GeV (see text for details).
}\label{tab:fieldsrc}
\centering
{\small
\begin{tabular}{crrccrccclll}
\hline
   MST cluster  &     RA   &   DEC       & $ l $  &  $ b $   & $n$ & $g$   & $M$    & $R_\mathrm{m}$ & $R_\mathrm{max}$ & $\Delta \theta$    & Notes \\ 
                &   J2000  &  J2000      &  deg   &  deg     &     &       &        &  deg  & deg   &   arcmin  &   \\
\hline
MST 0747$-$3313 & 116.8212 & $-$33.1776 & 248.4199 &  $-$4.0029 &  7 & 6.612 & 46.284 & 0.027 & 0.076 & 0.6 & 3FGL J0747.2$-$3311 \\
MST 0757$-$3728 & 119.3012 & $-$37.4768 & 253.1631 &  $-$4.4466 & 14 & 2.483 & 34.759 & 0.181 & 0.385 &     &   \\
MST 0954$-$5741 & 136.2318 & $-$57.6689 & 276.1837 &  $-$7.0924 & 13 & 3.683 & 47.880 & 0.089 & 0.224 & 3.9 & 3FGL J0904.8$-$5734  \\ 
\hline
MST 0757$-$3728 & 119.2832 & $-$37.4920 & 253.1687 &  $-$4.4668 & 22 & 3.180 & 69.950 & 0.084 & 0.342 &     &   \\
\hline
MST 0757$-$3728 & 119.2893 & $-$37.5216 & 253.1967 &  $-$4.4779 & 22 & 2.686 & 59.081 & 0.061 & 0.136 &     &   \\
\hline
\end{tabular}
}
\end{table*}

\section{New MST photon clusters}

As mentioned before, in our previous works we limited our analysis of the $\gamma$-ray sky at Galactic latitudes 
$|b| > 25\degr$ to avoid large background inhomogeneities.
In a subsequent study we started to extend MST cluster search to regions closer to the Galactic
belt, but considering higher selection thresholds to reduce the number of fake detections.
Two interesting clusters were found in these searches, and their main properties are described in the
following. 

\subsection{The cluster \ms}

We analyzed a $38\degr\times4\degr$ region bounded by the Galactic coordinates 
$242\degr < l < 280\degr$, $-12\degr < b < -4\degr$ in the 20--300 GeV energy range.
Adopting a severe selection length $\Lambda_\mathrm{cut} = 0.6\,\Lambda_m$ \citep[see][for details]{campana08,campana13}
and applying the further threshold $M \geq 25$ we found only three clusters, whose coordinates and MST parameters are
given in Table~\ref{tab:fieldsrc}. 
The coordinates of two clusters are very close to those of 3FGL sources within a few arcminutes,
while the third one, \ms, has been never reported before.
A refined search for this last source in a narrower strip ($-7\fdg5 < b < -3\fdg5$) 
at energies higher than 10~GeV gave a much more significant detection (see Table~\ref{tab:fieldsrc}) with only 
slight changes of the centroid coordinates but compatible with the previous ones.
A further analysis at $E>3$~GeV gave a cluster with 22 photons but again with a highly significant 
magnitude (last line in Table~\ref{tab:fieldsrc}) confirming the robustness of the detection.

False colour images of the photon density are given in the two upper panels of Figure~\ref{f:fig1}: note how for 
$E>10$~GeV (right panel) the photon cluster is readily apparent and its size is very similar to 
those of 3FGL sources, while for $E>3$~GeV (left panel) it is still detectable but the 
local background is much higher. 
No correspondence for this cluster was found in the 2nd LAT Pulsar Catalogue \citep{abdo13} 
and in the LAT Supernova Remnants Catalogue \citep{acero16}.

\subsection{The cluster \mfs}

MST analysis was also applied to events in the 10--300 GeV energy range in two belts with Galactic coordinates between
$14\degr < |b| < 26\degr$ and applying, as in the previous case, the threshold 
$M \geq 25$.
Of the 294 resulting clusters, 231 are in the 3FGL catalogue. Furthermore, after a more
robust filtering with $g > 3.5$ in order to reduce the possibility of selecting extended
features, the number of new clusters decreased to 23.
The one with the highest $M$ among them (centroid coordinates $RA = 296\fdg2141$ 
and $Dec = -21\fdg7190$; $l = 18\fdg5665$; $b = -21\fdg0826$)  has 20 photons 
and $M = 161.6$, much higher than in the other cases and typical of safely confirmed 
$\gamma$-ray sources.
We found that a source at a very close position (1\farcm9 from the cluster centroid coordinates) 
was already reported in the more recent 2FHL catalogue as \sr, but no analysis on its 
possible counterparts was presented. 

We performed a refined MST analysis in a small sky region $8\degr\times8\degr$
wide at different energies and dividing the entire data set into three equal time intervals,
each of length 28 months. 
The results using the two lower energy thresholds $E > 3$ GeV and $E > 10$ GeV are reported in Table~\ref{tab:M3int}, and 
clearly show that the source was brighter in the last interval (III).
During interval I the source is well detected at $E>3$~GeV, while using the $E>10$~GeV data the cluster has a lower significance.
In interval II the cluster 
significance was at the same level of background fluctuations for both values of the lower energy thresholds.
There is no doubt that it should also be highly variable and not detectable in the observation
periods covered by other \emph{Fermi}-LAT catalogues. 

\begin{table} 
\caption{Magnitude $M$ values of the cluster \mfss in three time intervals
and with the two lower energy thresholds $E>3$~GeV and $E>10$~GeV.}\label{tab:M3int}
\centering
{\small
\begin{tabular}{crr}
\hline
   Time interval &    $E > 3$ GeV  & $E > 10$ GeV  \\ 
\hline
      I          &    35.66        &   31.75       \\ 
      II         &    23.47        &    8.20       \\ 
      III        &    234.10        &  226.22       \\
\hline
\end{tabular}
}
\end{table}

\begin{table} 
\caption{Results of the ML analysis for the two clusters. The photon flux is in
units of 10$^{-10}$ ph cm$^{-2}$ s$^{-1}$.
}\label{tab:MLres}
\centering
{\small
\begin{tabular}{ccc}
\hline
                  &    \ms          &    \mfs        \\ 
\hline
   $TS$           &    51.76        &  185.53        \\
 Flux ($E>3$ GeV) &   $1.3\pm0.5$   &  $2.6\pm0.4$   \\
 Photon index     &   $1.8\pm0.6$   &  $2.2\pm0.2$  \\
\hline
\end{tabular}
}
\end{table}

Photon density maps are given in the lower panels of Figure~\ref{f:fig1}: note 
that for $E>10$~GeV (right panel) the photon cluster is very well apparent and that for 
$E>3$~GeV (left panel) it remains still detectable despite a higher local background. 

\section{ML analysis and light curves}

We verified the significance of both these clusters by means of the standard Maximum Likelihood 
(ML) analysis on the entire data set using the Science Tools package.
We used the \texttt{gtselect} tool to apply the cuts suggested by the 
\emph{Fermi}-LAT collaboration for point-like 
sources\footnote{\url{http://fermi.gsfc.nasa.gov/ssc/data/analysis/LAT_caveats.html}} 
to minimize the impact of the systematics and the contamination from non-photon events.  
The Region of Interest (ROI), with a radius of 10\degr, was centered at the MST cluster 
centroid coordinates and the photons were extracted with energies higher than 3 GeV applying the 
\texttt{P8SOURCE\_V6} event class selection and the corresponding Instrument Response Functions (IRF).
In the analysis we considered all the 3FGL sources within the ROI from the cluster centroid, 
as well as the Galactic and extragalactic diffuse emission. 
Furthermore, a source with a power-law spectral distribution was assumed at the MST coordinates.
The normalization of the 3FGL sources with $TS > 25$ within the ROI was allowed to vary in 
the fitting, while the parameters of the sources with $TS < 25$ were fixed to their 
catalogue values.  
The results are reported in Table~\ref{tab:MLres}.

\begin{figure}
\centering
\includegraphics[width=.5\textwidth]{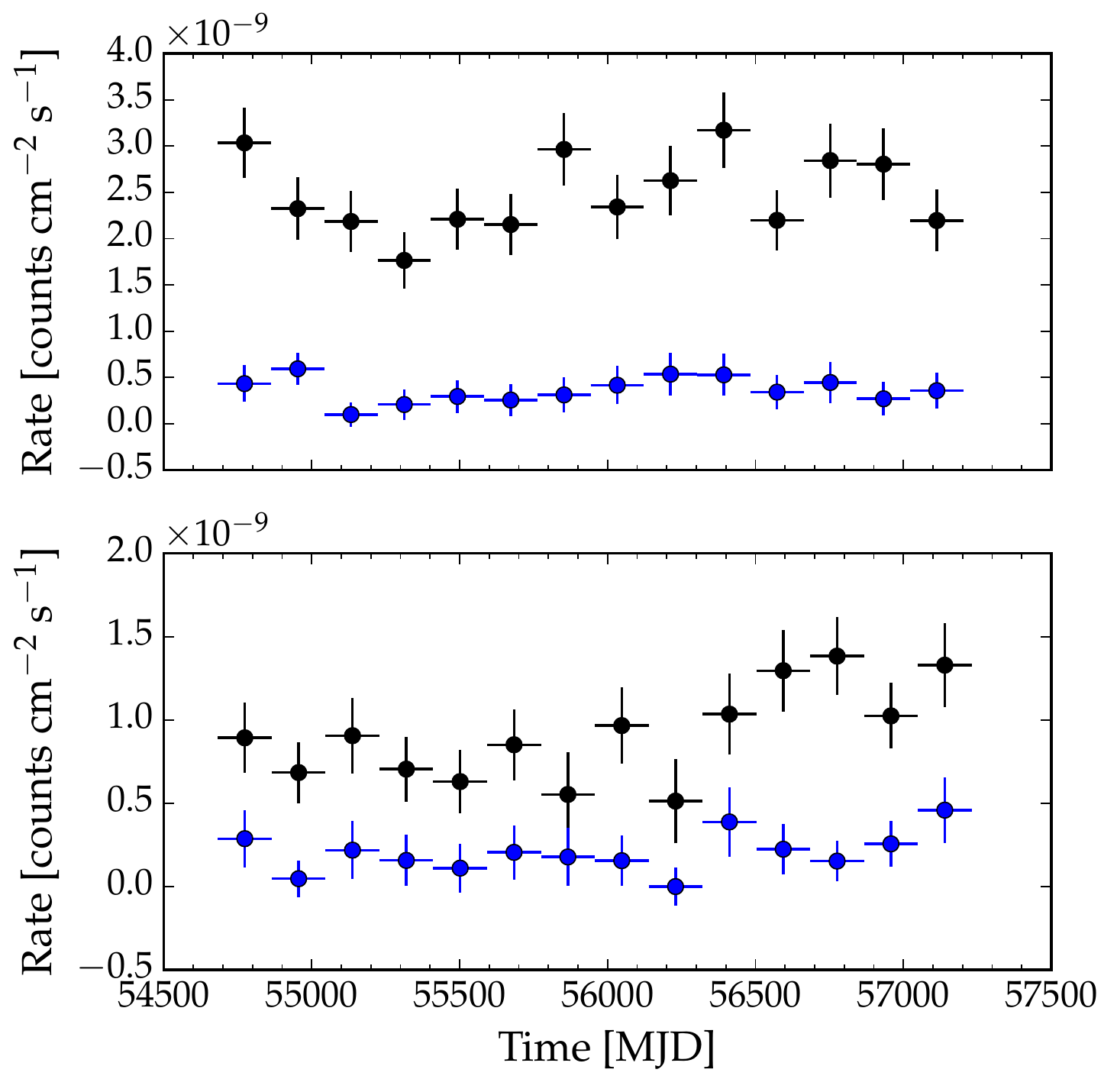}
\caption{Aperture photometry light curve of the photons at energy higher than 3 GeV (black points)
and 10 GeV (blue points) for \ms\ (upper panel) and \sr\ (lower panel).
Time binning is 6 months.
}
\label{f:LCsources}
\end{figure}

In addition, 
we performed a simple photometry following the standard \emph{Fermi} LAT Aperture Photometry 
procedure\footnote{\url{http://fermi.gsfc.gov/ssc/data/analysis/scitools/aperture_photometry.html}}
in order to inspect a possible variability in the long-term light curve (LC) of the source. 
Figure~\ref{f:LCsources} 
shows the resulting 6-month binned LC for the entire period (2008 Aug 04 -- 2014 Aug 04) 
in the 3--300 and 10--300 GeV energy bands, obtained selecting events within 1\degr\ aperture radius.
No model source has been assumed. 
In agreement with the MST analysis (Table~\ref{tab:M3int}), the \sr\ source was brighter in the last period, 
while for \ms\ a shallow minimum of emission could have occurred around MJD$\sim$55400.

\section{The search for possible counterparts}

\subsection{The cluster \ms}

Considering that the $\gamma$-ray source coordinates above 
10 GeV are given with an accuracy of a few arcminutes, we searched for possible counterparts within a 
cone of radius of 6\arcmin\ centred at the source position using the ASDC 
sky explorer tool\footnote{\url{http://www.asdc.asi.it/}}, that allows us 
to take into account many catalogues.

There are only two NVSS radio sources: one is quite weak (3 mJy) and without high confidence
counterparts in other bands. The much more interesting source NVSS~J075714$-$372734 
has a flux density of 79.2~mJy at 1.4~GHz.
In the second epoch Molonglo Galactic Plane Survey \citep{murphy07} at 0.843~GHz 
the corresponding source is MGPS J075714$-$372737, with a flux density of 99.3~mJy,
indicating that its radio spectrum is flat. Moreover, at the rather large angular separation 
of 1\farcm8 there is also a PMN source with a flux density of $88\pm10$~mJy at 4.85~GHz, comparable 
to the previous data. This, if actually associated with the NVSS source, would 
confirm the flat spectrum.

\begin{figure}
\centering
\includegraphics[width=.5\textwidth]{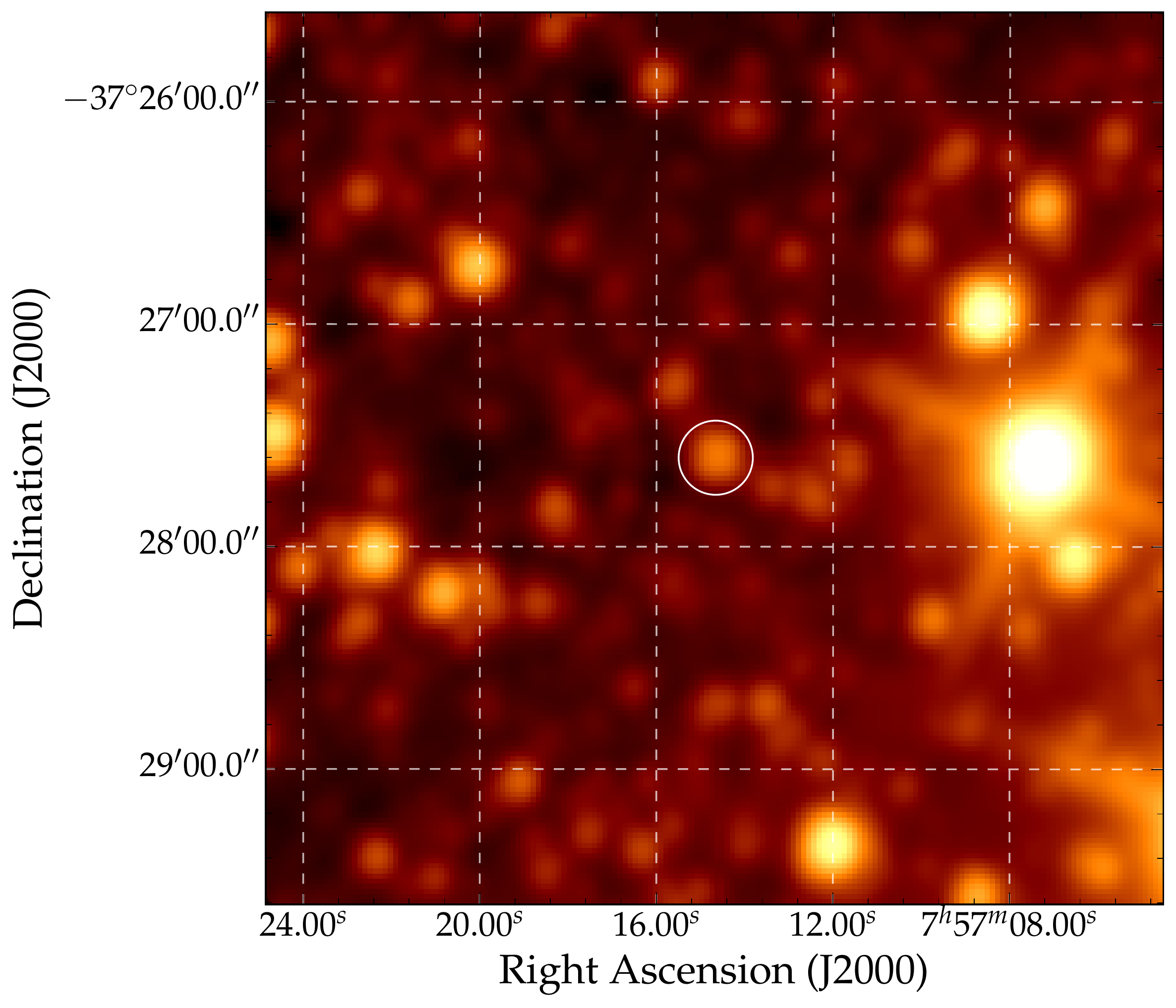}
\caption{
WISE image in the W1 ($[3.4]\mbox{ $\mu$m}$) bandpass with the likely infrared counterpart 
to the radio source NVSS~J075714$-$372734, indicated by the circle.
Image side is 4\arcmin.
}
\label{f:wise}
\end{figure}

At an angular distance of 2\farcs1, fully compatible with the NVSS resolution, 
there is a very faint optical object in the ESO DSS2-red image with very uncertain 
USNO photometric data.
This source, however, is positionally consistent with a brighter mid IR source 
AllWISE~J075714.66-372736.0
in the Wide-field Infrared Survey Explorer\footnote{\url{http://wise2.ipac.caltech.edu/docs/release/allsky/}}  
\citep[WISE;][]{wright10}, 
detected in all the four bandpasses with the following magnitudes: 
$[3.4]\mbox{ $\mu$m} = (13.435 \pm 0.026)$ mag,
$[4.6]\mbox{ $\mu$m} = (12.625 \pm 0.024)$ mag,
$[12]\mbox{ $\mu$m} = (10.303  \pm 0.059)$ mag,
$[22]\mbox{ $\mu$m} = (8.683  \pm 0.348)$ mag.
The WISE image in the $[3.4]\mbox{ $\mu$m}$ band is given in Figure~\ref{f:wise}.
In the [3.4]-[4.6]-[12] $\mu$m colour-colour diagram, these IR colours are consistent 
with those typical of a High-synchrotron-peaked BL Lac (HBL) as described in \cite{fmassaro11} 
for the TeV BL Lac objects. 
This source is also included in the WIBRALS catalogue 
\citep{dabrusco14} as a class C (\emph{candidate}) blazar.  
Optical spectroscopy of this source is not available and therefore we cannot safely establish 
whether it is a BL Lac object or a FSRQ, nevertheless its blazar nature appears well established 
and therefore it must be considered as the most interesting candidate for the counterpart 
of our new source.

\subsection{The cluster \mfs\ / \sr}

\begin{figure}
\centering
\includegraphics[width=.5\textwidth]{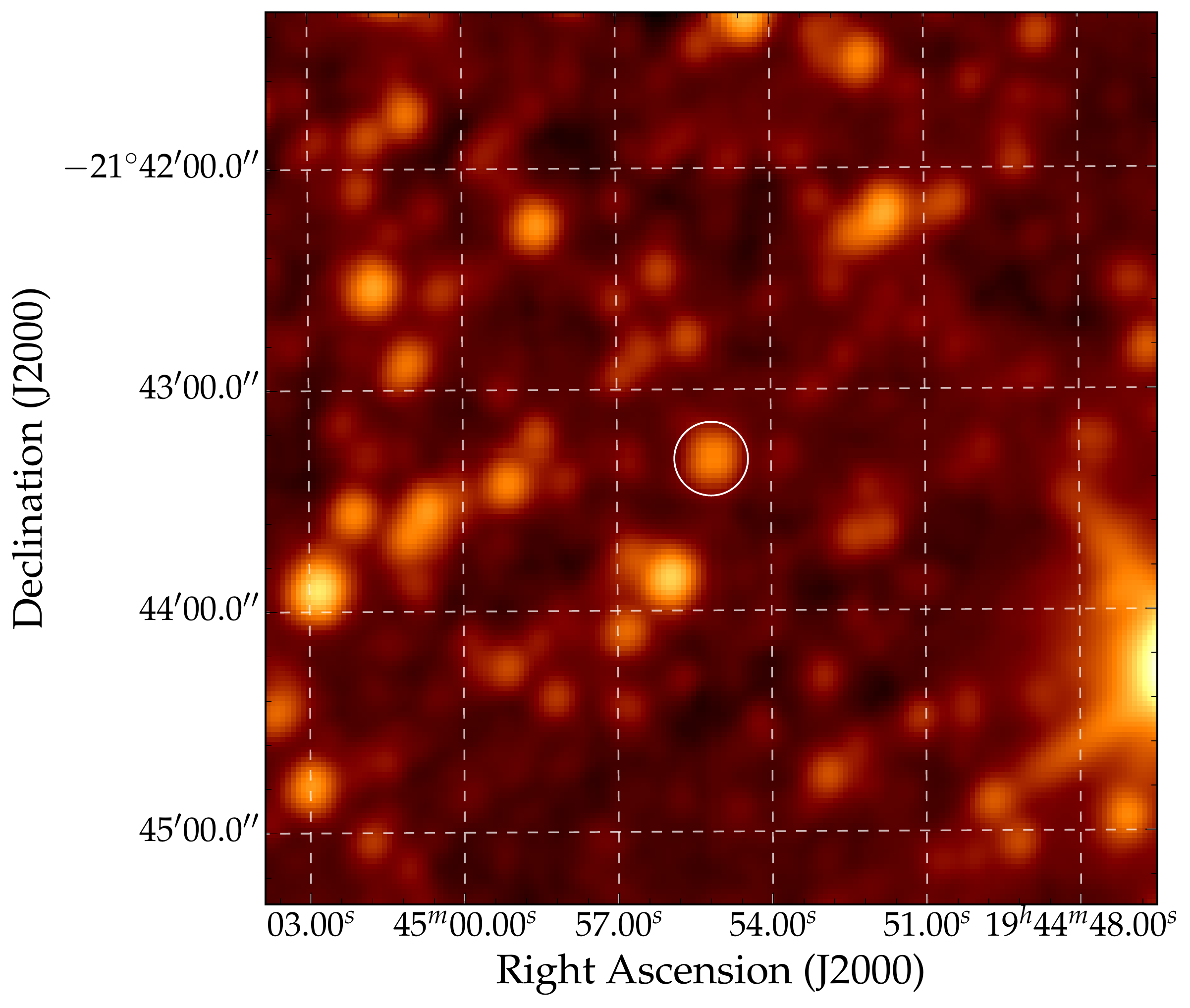}
\caption{
WISE image in the W1 bandpass ([3.4] $\mu$m) of the region with the likely counterpart to the radio 
source NVSS~J194455$-$214320, indicated by the circle. Image side is 4\arcmin.
}
\label{f:dss}
\end{figure}

In this case five NVSS sources are found close enough to the cluster position, but
the brightest and most interesting is NVSS~J194455$-$214320 (Figure~\ref{f:dss}) with a flux density at 
1.4 GHz of 42.1 mJy, likely associated with the close X-ray source 1RXS J194455.3$-$214318.
These sources are practically coincident with an optical object with GSC~2.2 magnitudes
$B = 18.41$ mag and $R = 17.93$ mag. 
We found only one other radio measurement of this source in the WISH catalogue \citep{debreuck02},
and its flux density at 0.352~GHz is 49 mJy, indicating a flat radio spectrum with a spectral index
$-0.12\pm0.05$, corresponding to a SED exponent of $0.88\pm0.05$:
for this reason it was included in the LORCAT catalogue \citep{fmassaro14}
aimed at the identification of potential counterparts of $\gamma$-ray sources likely
associated with blazars.

This source is also positionally well consistent with the IR counterpart 
AllWISE~J194455.16$-$214319.2, detected in all the four bandpasses with the following magnitudes: 
$[3.4]\mbox{ $\mu$m} = (13.970 \pm 0.028)$ mag,
$[4.6]\mbox{ $\mu$m} = (13.377 \pm 0.034)$ mag,
$[12]\mbox{ $\mu$m} =  (11.260  \pm 0.185)$ mag,
$[22]\mbox{ $\mu$m} =  (8.755  \pm 0.491)$ mag.
As for the previous cluster the corresponding point in the [3.4]-[4.6]-[12] \(\mu\)m 
colour-colour diagram is consistent with a High-synchrotron-peaked BL Lac (HBL) as 
described in \cite{fmassaro11} for the TeV BL Lac objects. 

\begin{figure}
\centering
\includegraphics[width=.5\textwidth]{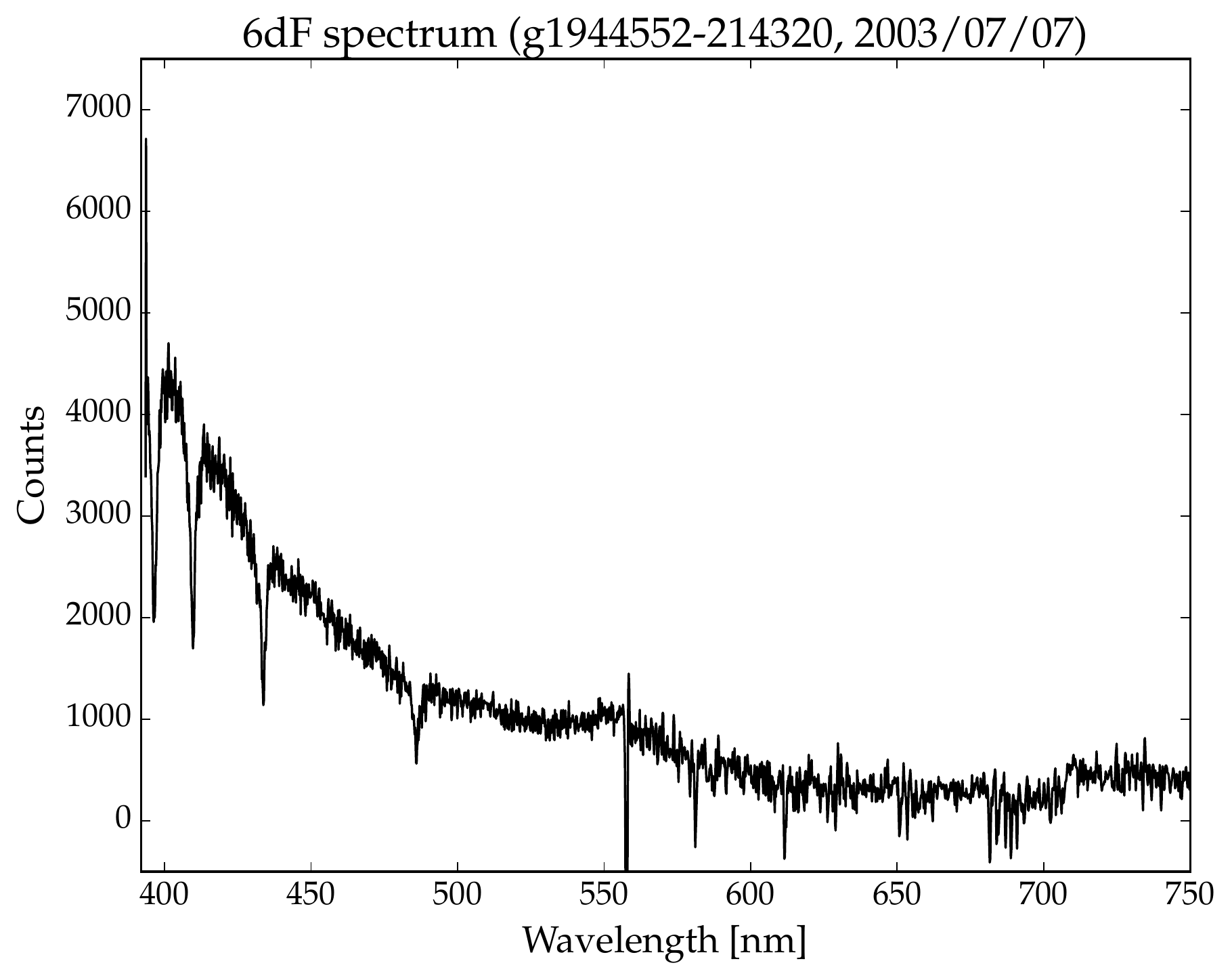}
\caption{
The 6dF spectrum of the possible optical counterpart to NVSS~J194455$-$214320:
note the absorption lines of the Balmer series. 
}
\label{f:6dF_spectrum}
\end{figure}

There is a good quality optical spectrum available from the 6dF database \citep{jones09},
here reported in Figure~\ref{f:6dF_spectrum}: surprisingly, it has a clear stellar appearance with well 
defined Balmer lines with the exception of the H$\alpha$, and it is classified as a 
DA white dwarf.

\begin{figure*}
\centering
\includegraphics[width=.49\textwidth]{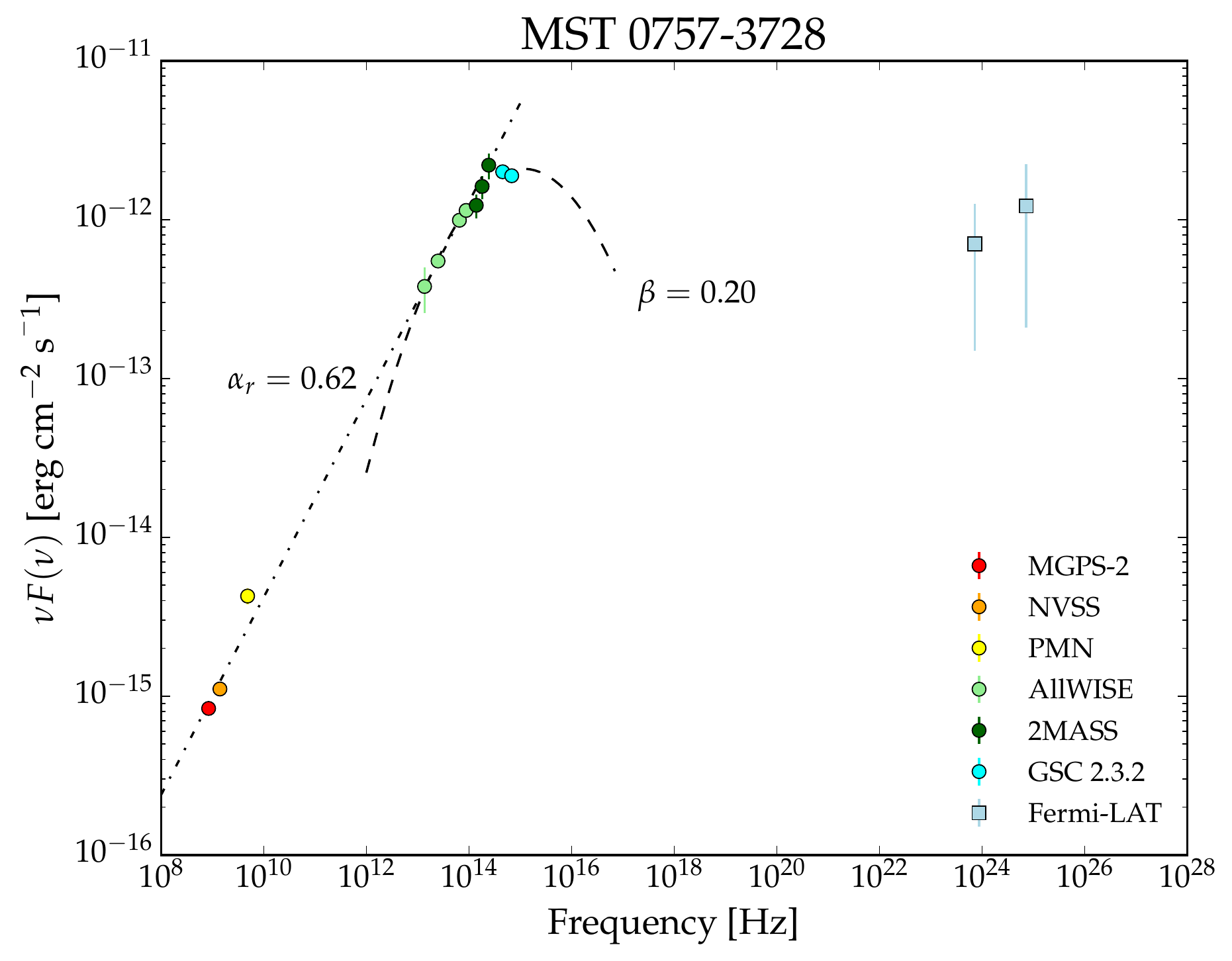}
\includegraphics[width=.49\textwidth]{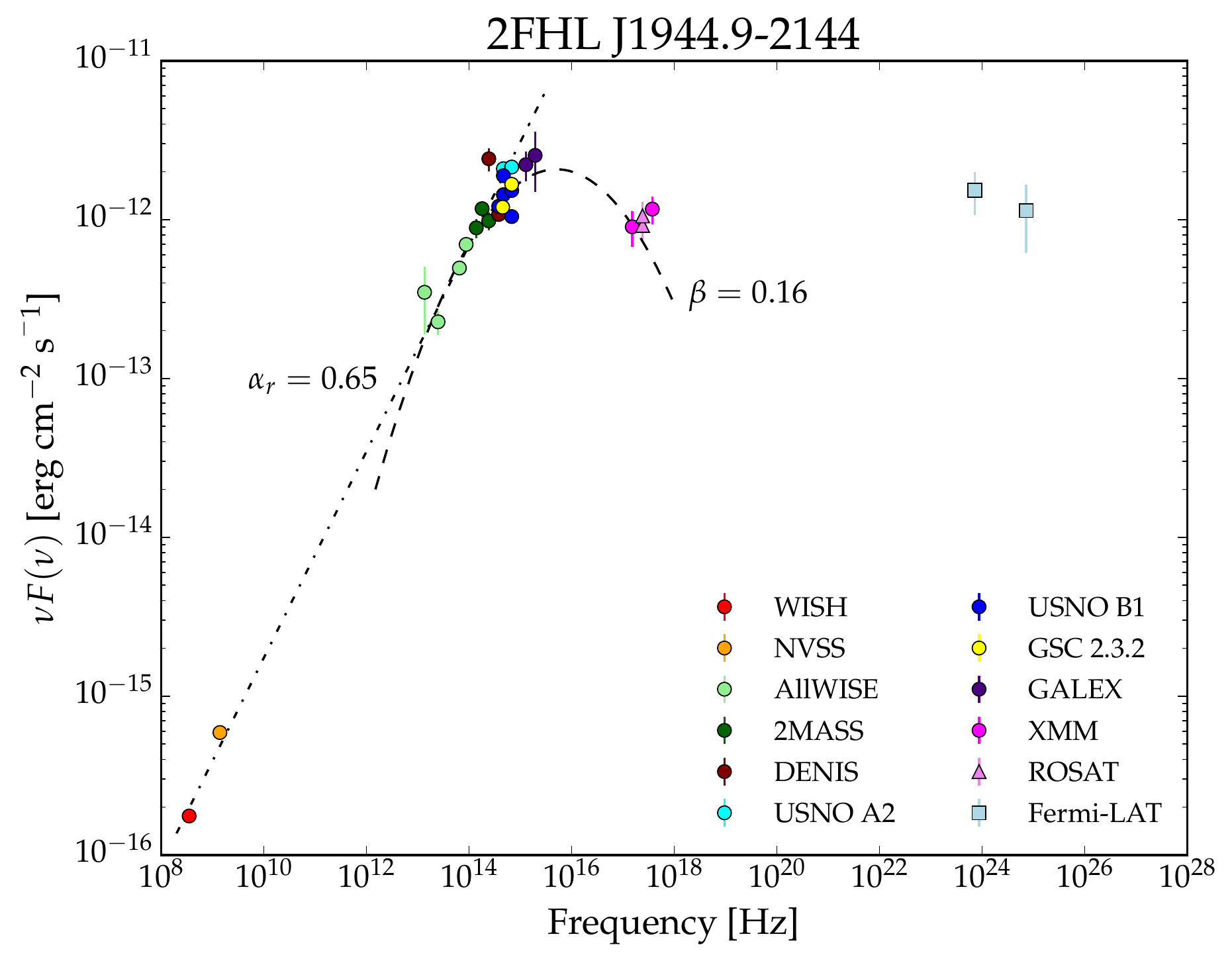}
\caption{The Spectral Energy Distribution of \ms\ (left panel) and of \sr\ (right panel).}
\label{f:sed}
\end{figure*}

\subsection{The Spectral Energy Distributions}

Using the good photometric data available from literature and databases we built the
Spectral Energy Distributions (SED) of both $\gamma$-ray sources and their possible counterparts.
The resulting plots are presented in the two panels of Figure~\ref{f:sed}: the two SEDs are remarkably
similar to that of a typical HSP source \citep[see, for instance,][]{giommi12}.
We fitted a log-parabola:
\begin{equation}
S(E) = A E^{-a-\beta\,\mathrm{Log}E}
\end{equation}
to data from the IR to the UV/X-ray range to estimate the frequency
of the
synchrotron bump and obtained $\approx$10$^{15}$ Hz and $\approx$$5\times 10^{15}$ Hz for
\mss and \sr, respectively.
Also, the curvature parameter $\beta$ resulted equal to 0.16 and 0.20, values frequently
found for blazars.
In Figure~\ref{f:sed} is shown also a linear fit $S(E)= KE^{-\alpha_r}$, carried out considering only the radio
data up to the frequencies corresponding to the AllWISE filters.
For \sr\ the different slope with respect to the one given by radio data alone is due to the constraint of matching the mid IR data.

\section{Summary and discussion}

The high energy $\gamma$-ray sky, according to the new scenario derived from 
\emph{Fermi}-LAT observations, appears characterised by the presence of several thousands of 
sources, many of which are variable over different time scales \citep{fmassaro16}.
The improvement in the data quality reached with the Pass 8 release and the 
continuous enrichment due to the extension of the \emph{Fermi}-LAT sky monitoring 
make possible the detection of new interesting sources.
In the present paper we report the discovery of two $\gamma$-ray 
sources with high significance: one in a region close to the Galactic equator, and the other having an
incongruous possible counterpart.   

The time and spectral behaviour of \mss in the $\gamma$-ray band appear blazar-like and therefore we are confident about the proposed counterpart; 
we cannot safely establish, however, whether it is a BL Lac object or a FSRQ because 
of the lack of optical spectroscopy.
Radio and $\gamma$-ray spectra are suggestive of a quasar, while the IR WISE 
colours are those of a HBL source.

The characteristics of the proposed counterpart to \sr, and particularly its optical 
spectrum, are quite unusual.
White dwarf DA stars, in fact, are neither known as flat spectrum radio sources nor as $\gamma$-ray 
emitters.
Cross-matching a sample of 14120 DA-type stars from the SDSS DR7 WD catalogue 
\citep{kleinman13}
with the 3FGL catalogue within a 6\arcmin\ angular distance, resulted in only 25 of them possibly associated with
$\gamma$-ray sources: 22 of which with a counterpart in the Roma-BZCAT 
\citep[5th Edition,][]{massaro14,massaro15}
blazar catalogue and the remaining three close to CRATES \citep{healey07}
radio sources and therefore likely associated with other blazars not yet optically 
identified.

The possibility that this object can be considered as a NS-WD binary, as in the case of the recently discovered 
millisecond pulsar (MSP) PSR J1824+10 in the error box of 3FGL J1824.0+1017 \citep{cromartie16}, 
cannot be excluded.
Even the occurrence of a  flat radio spectrum below 1 GHz does not rule out this possibility 
because, although rare, it is observed in a few MSPs \citep{kondratiev16}.
However, some alternatives must be taken into account: the possibility, although quite rare, 
that the close angular separation between a
blazar and a brighter WD might be due to chance, or the fact that the
reported 6dF spectrum does not correspond to the optical object.
We asked the 6dF data management to verify that the coordinates of their pointing
($RA = 19^\mathrm{h} 44^\mathrm{m} 55\fs16$, $Dec = -21\degr 43\arcmin 19\farcs5$; M.~Read,
private communication) were correctly reported, and obtained a confirmation.
The only way to disentangle this unusual finding is to perform new and more accurate
astrometric and spectroscopic observations of this puzzling object. 

\section*{Acknowledgements}
Part of this work is based on archival data, software or on-line services provided by the ASI Science Data Center (ASDC). 
We furthermore acknowledge use of archival Fermi-LAT data, of the Final Release 6dFGS archive, and of the SDSS and DSS archives.
We are particularly grateful to Mike A. Read of the 6dF team.

\bibliographystyle{mnras}
\bibliography{bibliography} 

\bsp	
\label{lastpage}
\end{document}